\documentclass[pra,twocolumn,amsfonts,superscriptaddress,a4paper,showpacs,showkeys]{revtex4}
\usepackage{graphicx} 
\begin{document}

\title{Critical Exponents \\of the
Two-layer Ising Model}\thanks{Supported by the National
Natural Science
Foundation of China under the project 19772074 and
by the Deutsche Forschungsgemeinschaft under the
project Schu 95/9-3.} 

\author{Z. B. Li}
\affiliation{Zhongshan University, Guangzhou 510275, China}
\affiliation{Associate Member of ICTP, Trieste, Italy}
\author{ Z. Shuai}
\affiliation{Universitat de Mons-Hainaut, 7000 Mons, Belgium}
\author{Q. Wang}
\author{H.J. Luo}
\author{L. Sch{\"u}lke}
\email{LSchuelke@t-online.de}
\affiliation{Universit{\"a}t Siegen, D-57068 Siegen, Germany\\~\\~\\}

\begin{abstract}
The symmetric two-layer Ising model (TLIM) is studied by the corner transfer matrix
renormalisation group method. The critical points and critical exponents are
calculated. It is found that the TLIM belongs to the same universality class as the
Ising model. The shift exponent is calculated to be 1.773, which is consistent with the
theoretical prediction 1.75 with 1.3\% deviation.
\end{abstract}



\maketitle

\paragraph{\bf Introduction}
The two-layer Ising model (TLIM), as a simple generalisation of the
two-dimensional Ising model and a simple model for the magnetic ultra-thin
films, has been studied for a long time \cite{Ball64, Alla70, Abe70,
Suzu71,Bind74, Oitm75}. Cobalt films grown on a Cu(100) crystal, for
instance, have highly anisotropic magnetisation \cite{Oepe90} and
could therefore 
be viewed as layered Ising-spin systems. It has been found that capping PtCo
in TbFeCo to form a two-layer structure has applicable features, for
instance, raising the Curie temperature and reducing the switching fields
for over-writable magneto-optical disks \cite{Shim92}. Apart from various
possible applications to real physical materials, the model is theoretically
interesting for its rich phase structure. The model has several interesting
equivalents, such as a two-species gauge invariant Ising model \cite{Li94}, a
spin-${\frac{3}{2}}$ Ising model \cite{Hori96}, a model of the dilute lattice
gas and a quantum spin-${\frac{1}{2}}$ ladder at zero temperature. The TLIM
is important for the investigation of crossover from the 2-dimensional Ising
model to the 3-dimensional one. In particular, it has been argued that the
critical point of the latter could be found from the spectrum of the TLIM
  \cite{Wosi94}.

In recent years, some approximation methods have been applied to this model 
\cite{Hori96,Ange92,Ange95, Ange97, Lipo93, Lipo98, Hori97,Hu99}. A
critical line has been found in all these studies. As expected, the Curie
temperature is very sensitive to the inter-layer interaction. Many
discussions have been directed to the shift exponent at the decoupling point.
Abe \cite{Abe70} and Suzuki 
\cite{Suzu71}  have predicted  $\gamma =7/4$ for the isotropic model many years ago. 
Only in recent years the
computational results have converged to the prediction with still about 
$2.3\%$ deviation \cite{Hori97, Lipo98}.

Apart from the shift exponent, it is also interesting to study the critical
behaviour along the critical line. The model has the same critical exponents
at the two ends of the critical line corresponding to the solvable decoupling
limit and the infinite interlayer coupling limit. But it is clear that the
decoupled system has a higher
 symmetry than the coupling layers, hence one cannot assume a priori
 that the TLIM belongs to the one-layer Ising
universality class. Unusual finite-size effects have been observed in the
dynamic process \cite{Li00}. It has been proposed that the critical exponents
(or some of them) would vary continuously along the critical line \cite{Li94,
Hori97}. However, there are also arguments in favour of unchanged
exponents. Based on their correlation length equality method, Angelini et al.
argued that the exponent $\nu $ is a constant \cite{Ange95}. The accurate
computation of critical exponents along the critical line remains  a
difficult unsolved problem.

It is our purpose to provide a reliable prediction for the critical exponents
based on the corner transfer matrix renormalisation group (CTMRG) method \cite
{Nish96a, Nish96b, Nish97}.

The CTMRG method follows White's idea of density matrix renormalisation
group  \cite{Whit92, Whit93}. It reduces the phase-space dimensions by
omitting the eigenstates corresponding to small eigenvalues of the density
matrix. But instead of constructing the density matrix through a series of
column transfer matrices, Nishino and Okunishi made use of the corner
transfer matrices. The CTMRG method has shown many merits for the
two-dimensional classical models which can be mapped into
interaction-around-face (IRF) models. Since the system can be easily extended
to very large lattices, 
the finite size effect can be ignored. The error mainly comes
from the finite dimensions of the renormalised phase space, which only
becomes severe for huge lattices at or very close to the critical point.
Numerically it has been observed that the extrapolation from the finite size
scaling can provide very accurate information about the criticality.

We will focus on the TLIM composed of two identical layers. In this case, the model
has a symmetry of interchanging the spins of the 
two layers. This symmetry is vital
to the critical behaviour of the model. A 
breakdown of universality due to a small symmetry breaking \cite{Hori97}
has been observed.
In order to keep this symmetry, special care should be taken in the
programming. Technically, the symmetry reduces the dimensions of the
transfer matrix remarkably. But the extra degeneracy will slow down the
convergence of the iteration which makes the problem much 
more difficult compared to the
one-layer Ising model.

The critical behaviour can be easily observed in various
quantities, such as the rapid increase of the magnetisation, the lambda
peak  of the capacity, and the anomalous slowing down of convergence of the
renormalisation group iterations. However, to calculate the critical
exponents, one needs to locate the critical points to high precision. This is
done by a careful  extrapolation of  the renormalisation dimension $m$.

In Section 2, the TLIM is defined and transformed into an IRF model which is
suitable for the CTMRG study. In the same section, the 
CTMRG method particular
for the considered model will be introduced. The finite-size scaling and
the finite renormalisation dimension effect is studied in Section 3. At the
same time the critical points are located, critical exponents of the scaling law
are computed. Results for the critical exponents $\eta $  and $\nu $ can be
found also in Section 3. Discussions and conclusions are given in the last
section.

\vspace{1cm}

\paragraph{\bf The CTMRG method for the TLIM}
The Hamiltonian of the model is defined by 
\begin{widetext}
\begin{equation}  \label{hami}
H=-J_{1}\sum_{<i,j>}s_{i}s_{j}-J_{2}\sum_{<i,j>}\sigma_{i}\sigma_{j}-
\lambda \sum_{i} s_{i}\sigma_{i}\;,
\end{equation}
\end{widetext}
where $s_{i}=\pm 1$ and $\sigma_{i}=\pm 1$ are Ising spins on two identical
square lattices. For concreteness, let us say, $s_{i}$ is at the site $i$ of
 layer 1, $\sigma_{i}$ is at the corresponding site of layer 2 which
is also labelled by $i$. Then $<i,j>$ should be understood as a nearest
neighbour pair of sites either on layer 1 or layer 2, depending on which of the
spin variable, $s$ or $\sigma$, is concerned. In this paper, we only
discuss the symmetric ferromagnetic case with nearest neighbour coupling 
$J_{1}=J_{2}=J >0$ and with interlayer coupling $\lambda=\rho J>0$.

To obtain the equivalent IRF model, two species of Ising spins, $u_{ij}=\pm
1 $ and $v_{ij}=\pm 1$, are introduced for each link $<i,j>$. The
Hamiltonian (\ref{hami}) is generalised to 
\begin{widetext}
\begin{equation}
H^{\prime }=-J^{\prime }\left[
\sum_{<i,j>}u_{ij}(s_{i}+s_{j})+\sum_{<i,j>}v_{ij}(\sigma _{i}+\sigma
_{j})+\rho ^{\prime }\sum_{i}s_{i}\sigma _{i}\right]  \label{irfh}\;,
\end{equation}
\end{widetext}
where $J^{\prime }$ and $\rho ^{\prime }$ are defined as 
\begin{eqnarray}
J^{\prime } &=&{\frac{1}{2}}\ln (e^{2J}+\sqrt{e^{4J}-1}),  \label{jrho} \\
\rho ^{\prime } &=&{\frac{\rho J}{J^{\prime }}}.
\end{eqnarray}
It is not difficult to show that, by summing over all $u-$ 
and $v-$spins, the TLIM is
recovered apart from a overall irrelevant constant factor to the partition
function. Inversely, if the s-spins and $\sigma $-spins are summed, one will
obtain the IRF model with four u-spin interaction and four v-spin
interaction around each vertex. The vertex weight of the IRF model is given
by 
\begin{widetext}
\begin{equation}
W_{abcd}=\sum_{s,\sigma =\pm 1}\exp\left\{ J^{\prime }\left[
s(u_{a}+u_{b}+u_{c}+u_{d})+\sigma (v_{a}+v_{b}+v_{c}+v_{d})+\rho ^{\prime
}s\sigma \right] \right\}  \label{verw}\;,
\end{equation}
\end{widetext}
where $a,b,c$, and $d$ denote the four (u,v)-spin pairs in the links joining
to the vertex. Special weights in the boundary, which have links less than
four, can be defined in a similar way as (\ref{verw}) but the values of $s$
and $\sigma $ are subject to the boundary condition. For instance, for the
fixed boundary condition, the weight along the boundary, $W_{abc}$, is defined
by 
\begin{equation}
W_{abc}=\exp \left[ J^{\prime }\left(
u_{a}+u_{b}+u_{c}+v_{a}+v_{b}+v_{c}+\rho ^{\prime }\right) \right]\;.
\label{bouw}
\end{equation}
The vertex at the corner has a weight 
\begin{equation}
W_{ab}=\exp \left[ J^{\prime }\left( u_{a}+u_{b}+v_{a}+v_{b}+\rho ^{\prime
}\right) \right]  \label{corw}\;. 
\end{equation}

\begin{figure}[htbp!]\centering  
\includegraphics*[width=8cm]{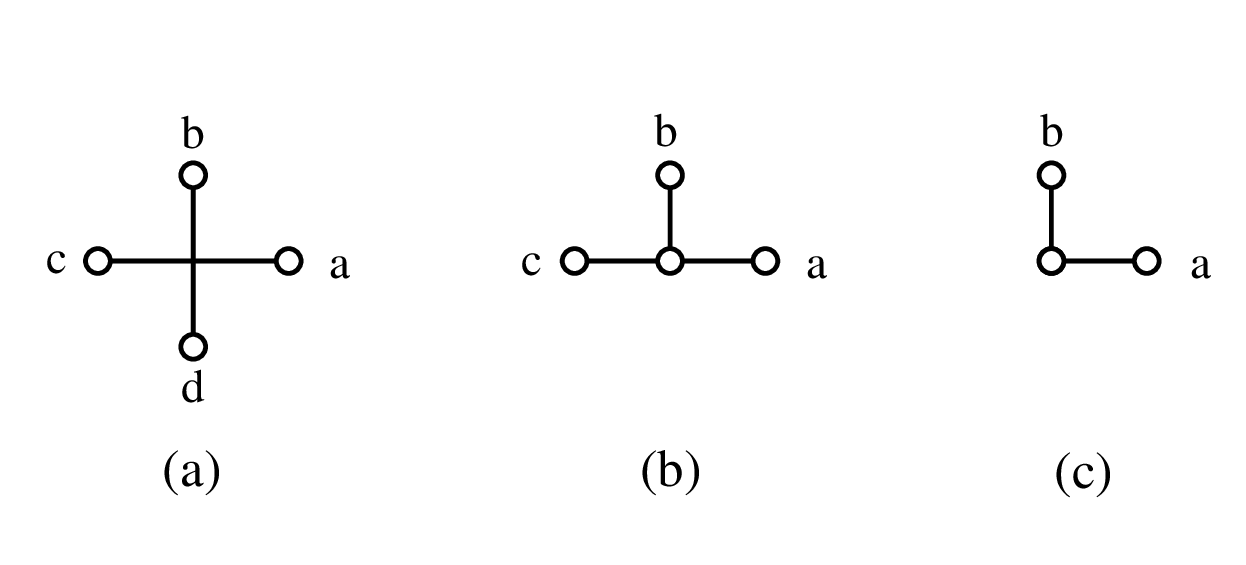} 
\caption{(a) A full vertex weight $W_{abcd}$; (b) an edge vertex weigh 
$W_{abc}$; (c) a corner vertex weigh $W_{ab}$.
}
\label{f1qw_k}
\end{figure}

They are schematically given in Figure~\ref{f1qw_k}a,b,c, respectively.

\begin{figure}[htbp!]\centering  
\includegraphics*[width=7cm]{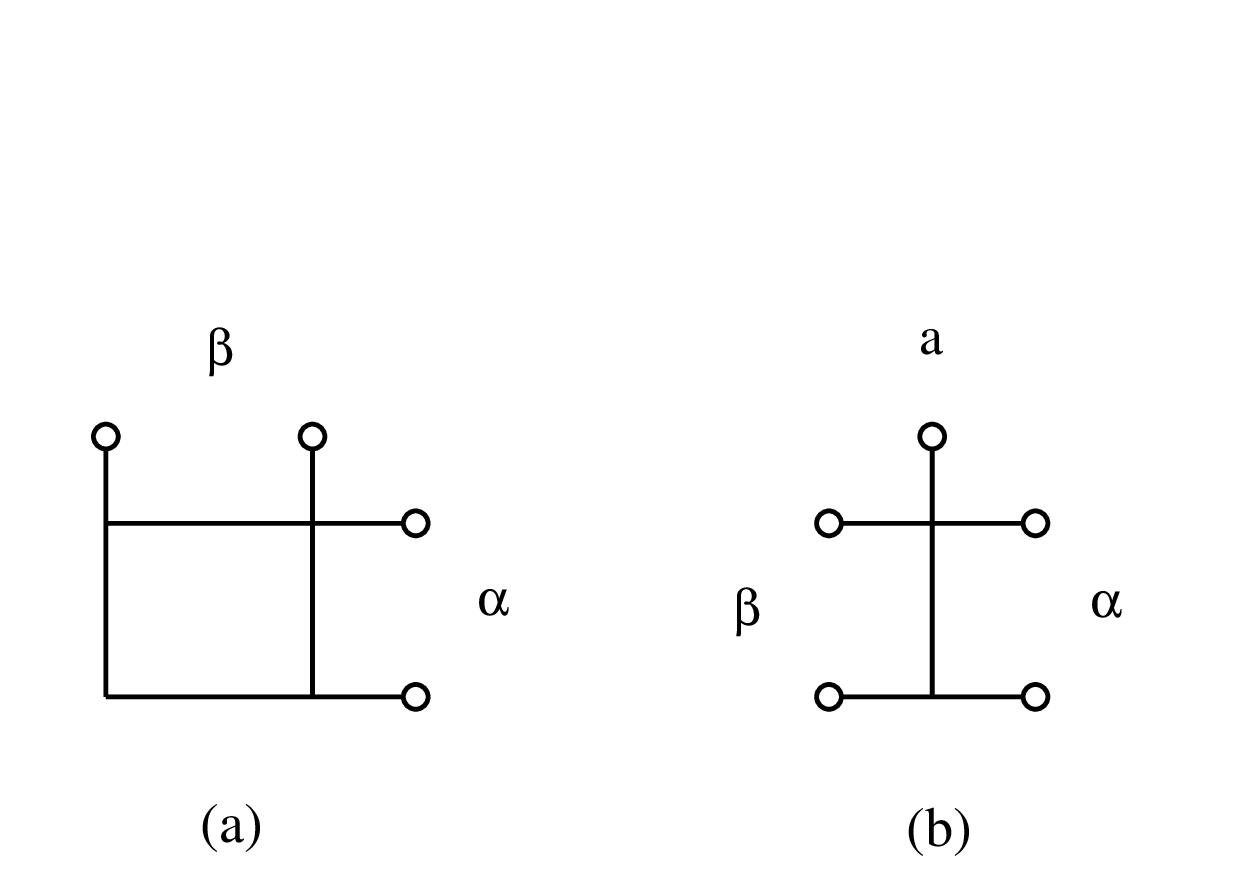} 
\caption{
(a) A corner transfer matrix (CTM), $C_{\alpha \beta }(2)$; a
half-row transfer matrix (HRTM), $P_{\alpha \beta ,a}(2)$.
}
\label{f2qw}
\end{figure}

The partition function is a trace of products of all vertex weights. The
trace here means to sum over all spins. In the following, whenever two
vertices are connected by a link, a summation over the (u,v)-spin pair in
that link is implied. The vertex weights of a quarter lattice, after
summation over all inner spins and boundary spins as it has been implied, is
called a corner transfer matrix(CTM), denoted by $C_{\alpha \beta }(N)$.
Here $\alpha $ and $\beta $ denote the spin configurations of two open
sides of the corner respectively, and
$N$ is the size of the corner. A schematic
representation for a CTM is given in Figure~\ref{f2qw}a.
\vspace*{-0.5cm}

Another matrix we need is the so-called half-row transfer matrix (HRTM)
denoted by $P_{\alpha\beta,a}(N)$. It is defined by the recursion relation 
\begin{eqnarray}  \label{hrtm}
P_{\alpha\beta,a}(N)&=&\sum_{c}W_{abcd}P_{\alpha^{\prime}
\beta^{\prime},c}(N-1)\;, \\
P_{bc, a}(1)&=&W_{abc}\;,
\end{eqnarray}
where $\alpha^{\prime}$ and $\beta^{\prime}$ are spin configurations of two
sides of the HRTM with size of $(N-1)$
; $c$ is the (u,v)-spin pair of the top of that
HRTM; $\alpha$ and $\beta$ are spin configurations of the enlarged HRTM with 
$\alpha$ representing $(\alpha^{\prime}, b)$ and $\beta$ representing 
$(\beta^{\prime}, d)$ respectively. For each value of $a$, 
$P_{\alpha\beta,a}(N)$ can also be considered as a matrix, denoted by 
$P_{a}(N)$. Figure~\ref{f2qw}b is an example of the HRTM.

\begin{figure}[htnp!]\centering  
\includegraphics*[width=7cm]{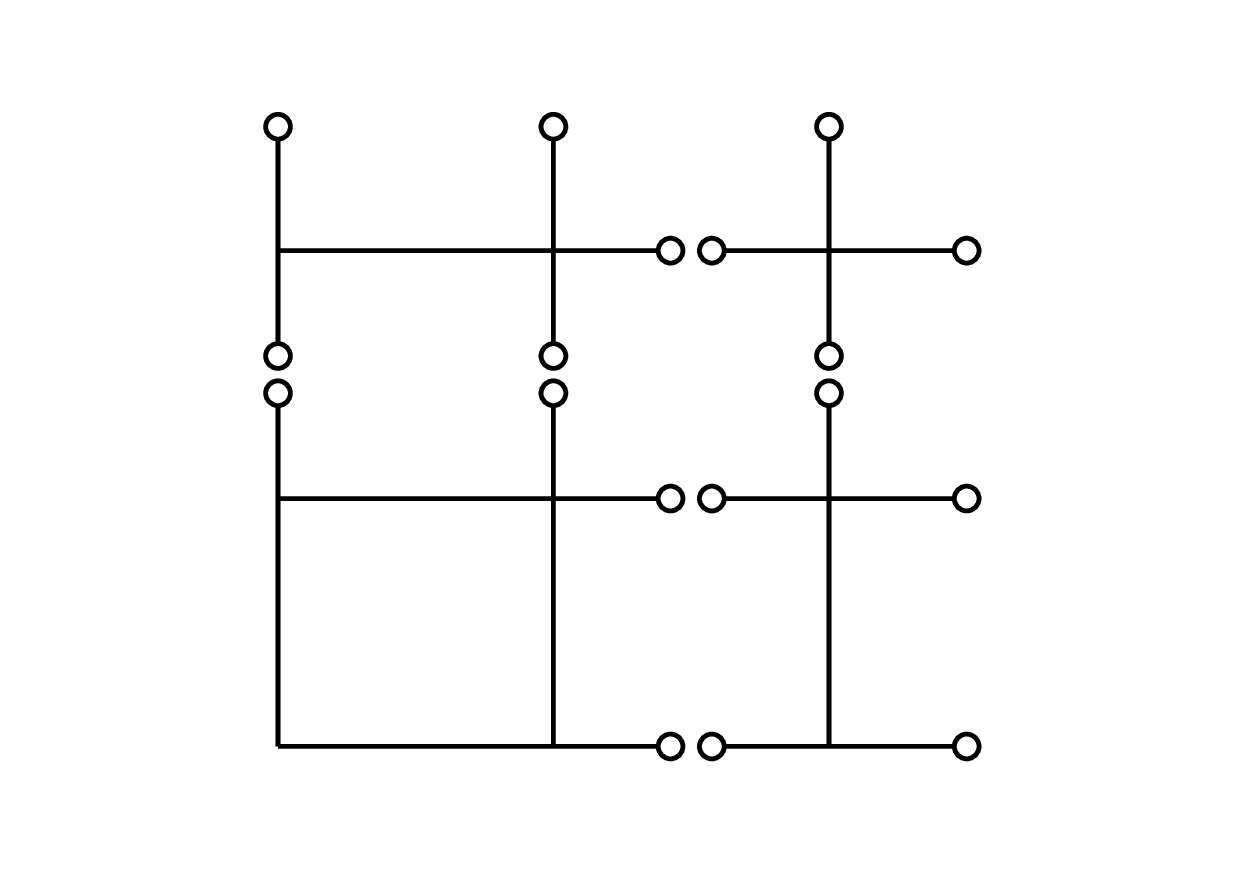} 
\caption{
Enlarging $C_{\alpha \beta }(2)$ into $C_{\alpha \beta }(3)$.
}
\label{f3qw}
\end{figure}

In the process of the CTMRG iterations, the size of CTM is enlarged 
by one lattice
spacing each time. This is done in this way: (1) glue a HRTM to each open
side of the CTM, (2) complement the bigger corner by adding a vertex weight.
The recursion relation is 
\begin{widetext}
\begin{equation}  \label{enla}
C_{\alpha\beta}(N)=\sum_{\alpha^{\prime\prime}\beta^{\prime\prime}}
\sum_{cd}W_{abcd}P_{\alpha^{\prime} \beta^{\prime\prime},c}(N-1)\;.
C_{\alpha^{\prime\prime}\beta^{\prime\prime}} (N-1) P_{\alpha^{\prime\prime}
\beta^{\prime},d}(N-1)\;.
\end{equation}
\end{widetext}
It is schematically shown in Figure~\ref{f3qw}.

By use of CTM, the partition function of an even size lattice is simply the
trace of the CTM to power four 
\begin{equation}
Z_{2N}=Tr(C(N)^{4}) . \label{part}
\end{equation}
Therefore, the CTM and the density matrix have common eigenvectors. Denote the
eigenvalues of $C$ by $\{\omega _{k}|k=0,1,2,...\}$, assuming that $\omega
_{0}\ge\omega _{1}\ge\omega _{2}\ge...$, the eigenvalues of the density matrix
are $\{\omega _{k}^{4}|k=0,1,2,...\}$. According the spirit of DMRG, only a
certain number of the biggest eigenstates will be kept. For instance, the
first $m$ eigenstates are kept, the eigenstates with smaller eigenvalues 
$\{\omega _{k}|k>m\}$ are ignored. $m$ is the so-called renormalisation
dimension.

The CTMRG iterations contains the following steps: (1) construct $C(N)$ of a
small CTM which is small enough to be exactly diagonalised; (2) construct 
$P_{a}(N)$; (3) diagonalise $C(N)$; (4) if the dimension of $C(N)$ is smaller
than $m$, all eigenstates are used as basis of the renormalised phase space,
otherwise only $m$ eigenstates with the biggest
eigenvalues are used as basis; (5) by use of the eigenvectors which have
been chosen for the renormalised space, form a projection
 operator $U(N)$ that
projects old spin states into the renormalised space; (6) use $U(N)$ to
project $P_{a}(N)$ to the renormalised HRTM $P_{a}^{r}(N)$  ; (7)
form a diagonal matrix $C^{r}(N)$ by the first $m$ largest eigenvalues,
which is the renormalised CTM; (8) use $P_{a}^{r}(N)$, $C^{r}(N)$, and the
vertex weight $W$ to form a bigger CTM, $C(N+1),$ and a bigger HRTM, 
$P(N+1)$, 
and repeat the above procedure from step (3) with $N$ replaced by $N+1$.

Since the renormalised phase space has dimension  $\le m$, the CTMRG
in principle can infinitely go on. Therefore the lattice can be easily
enlarged to very big sizes.

Particularly for the TLIM, it is efficient to choose the basis to be
symmetric and antisymmetric in the exchange of two layers. In this way, the
CTM is automatically block diagonal. The symmetric block and the
antisymmetric block can be diagonalised separately. This is not just
allowing larger $m$ and high precision, but  is also essential for keeping
the presumed symmetry between two layers.

For small interlayer coupling, there are many approximate degenerate
eigenstates due to the layer symmetry. If one does not separately consider
the symmetric and antisymmetric phase space, it is easy to destroy the layer
symmetry in the renormalisation procedure.

Following the method of Nishino and Okunishi \cite{Nish96a}, the
magnetisation $M$ can be calculated on an odd size lattice which is
constructed by inserting a HRTM matrix between two adjacent CTM matrices and
adding a proper vertex at the lattice centre. The probability  that the
central site has a spin state $(s,\sigma )$ is given by 
\begin{equation}
M^{s\sigma }={\frac{\sum_{abcd}X_{abcd}^{s\sigma }Tr\left(
P_{a}^{r}C^{r}P_{b}^{r}C^{r}P_{c}^{r}C^{r}P_{d}^{r}C^{r}\right) }{
\sum_{abcd}W_{abcd}Tr\left(
P_{a}^{r}C^{r}P_{b}^{r}C^{r}P_{c}^{r}C^{r}P_{d}^{r}C^{r}\right) }} \;,
\label{magnp}
\end{equation}
where $X_{abcd}^{s\sigma }$ is defined by 
\begin{widetext}
\begin{equation}
X_{abcd}^{s\sigma }=\exp\left\{ J^{\prime }\left[
s(u_{a}+u_{b}+u_{c}+u_{d})+\sigma (v_{a}+v_{b}+v_{c}+v_{d})+\rho ^{\prime
}s\sigma \right] \right\} \;. \label{magw}
\end{equation}
\end{widetext}
The magnetisation of layer 1, $M=<s>$, is 
\begin{equation}
M=M^{++}+M^{+-}-M^{--}-M^{-+} \;. \label{magn}
\end{equation}
Due to the layer symmetry, it is also one half of the total magnetisation.
The local energy density $E$ can be calculated in a similar way.
\begin{equation}
E={\frac{\sum_{abcd}Y_{abcd}Tr\left(
P_{a}^{r}C^{r}P_{b}^{r}C^{r}P_{c}^{r}C^{r}P_{d}^{r}C^{r}\right) }{
\sum_{abcd}W_{abcd}Tr\left(
P_{a}^{r}C^{r}P_{b}^{r}C^{r}P_{c}^{r}C^{r}P_{d}^{r}C^{r}\right) }} \;,
\label{ene}
\end{equation}
where
\begin{equation}
Y_{abcd}=\sum_{s,\sigma}X_{abcd}^{s\sigma }\ln (X_{abcd}^{s\sigma })\;.
\end{equation}

\vspace{1cm}

\paragraph{\bf Extrapolation and finite-size scaling}
At the critical point $J_{c}$, it is expected that $M$ obeys the finite-size
scaling form $M \sim L^{-(d-2+\eta )/2}$, where $L$ is the lattice size, so
\begin{equation}
\ln M(L)=const-\frac{d-2+\eta }{2}\ln L\;.
\end{equation}

If we calculate $\partial \ln M/\partial \ln L$ and plot it with $\ln L$
as the horizontal axis, the curve should be a horizontal line for the critical
point $J_{c}$ with the vertical value of the line as $-(d-2+\eta )/2$. For $J$
near $J_{c}$, there should be deviations to the power-law behaviour. A
simple physics consideration reveals that
\begin{figure}[htbp!]\centering  
\includegraphics*[width=7cm]{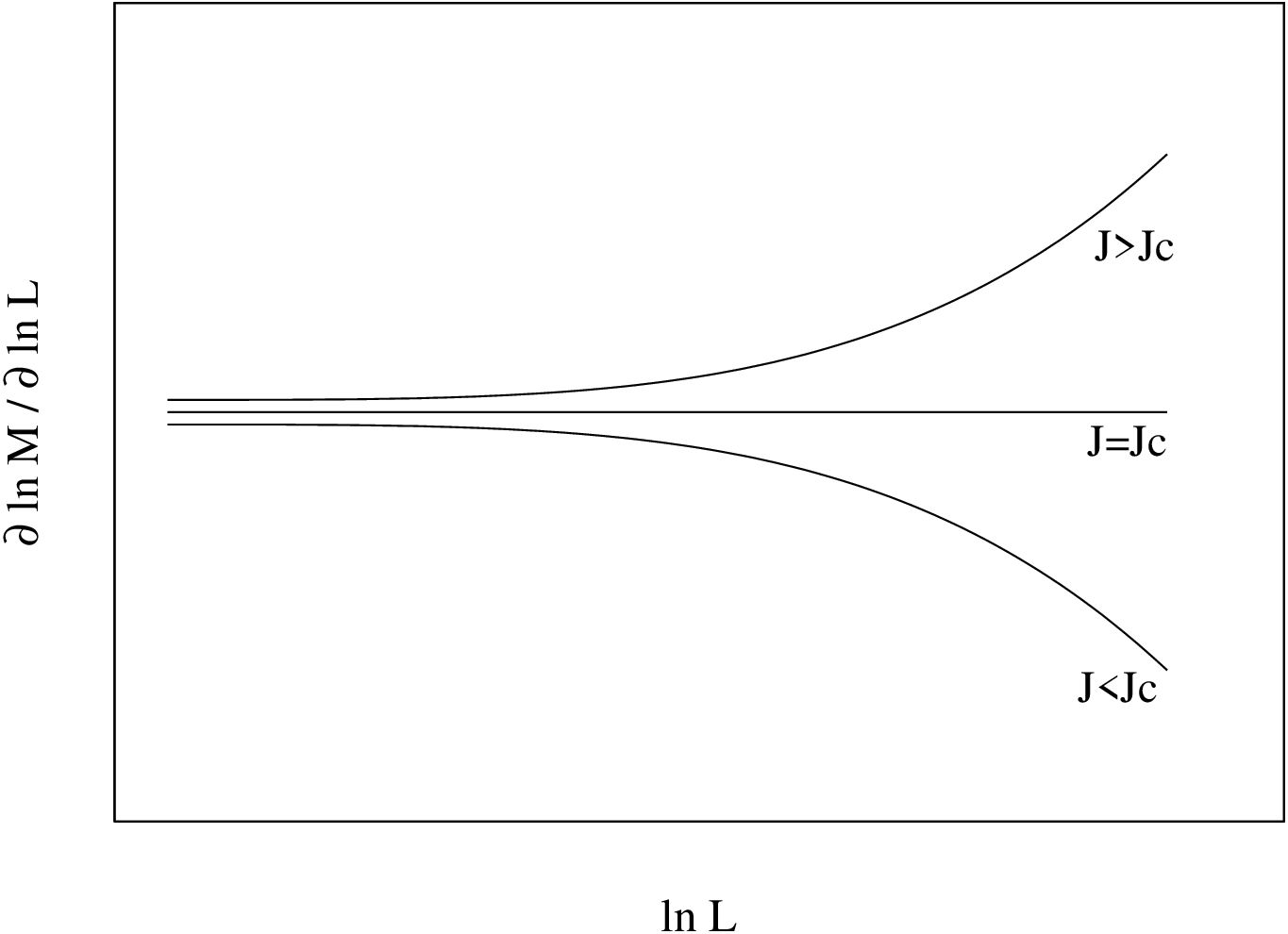} 
\caption{
Schematic plot of the ideal curves $\partial \ln M/\partial \ln L$ versus
$\ln L$.
}
\label{fig4}
\end{figure}
 the curves should be in the form as
shown in Figure~\ref{fig4}. 
So by searching for the ``best'' horizontal line, one can
determine the critical point $J_{c}$ as well as the critical exponent $\eta$.

\begin{figure}[htbp!]\centering  
\includegraphics*[width=7cm]{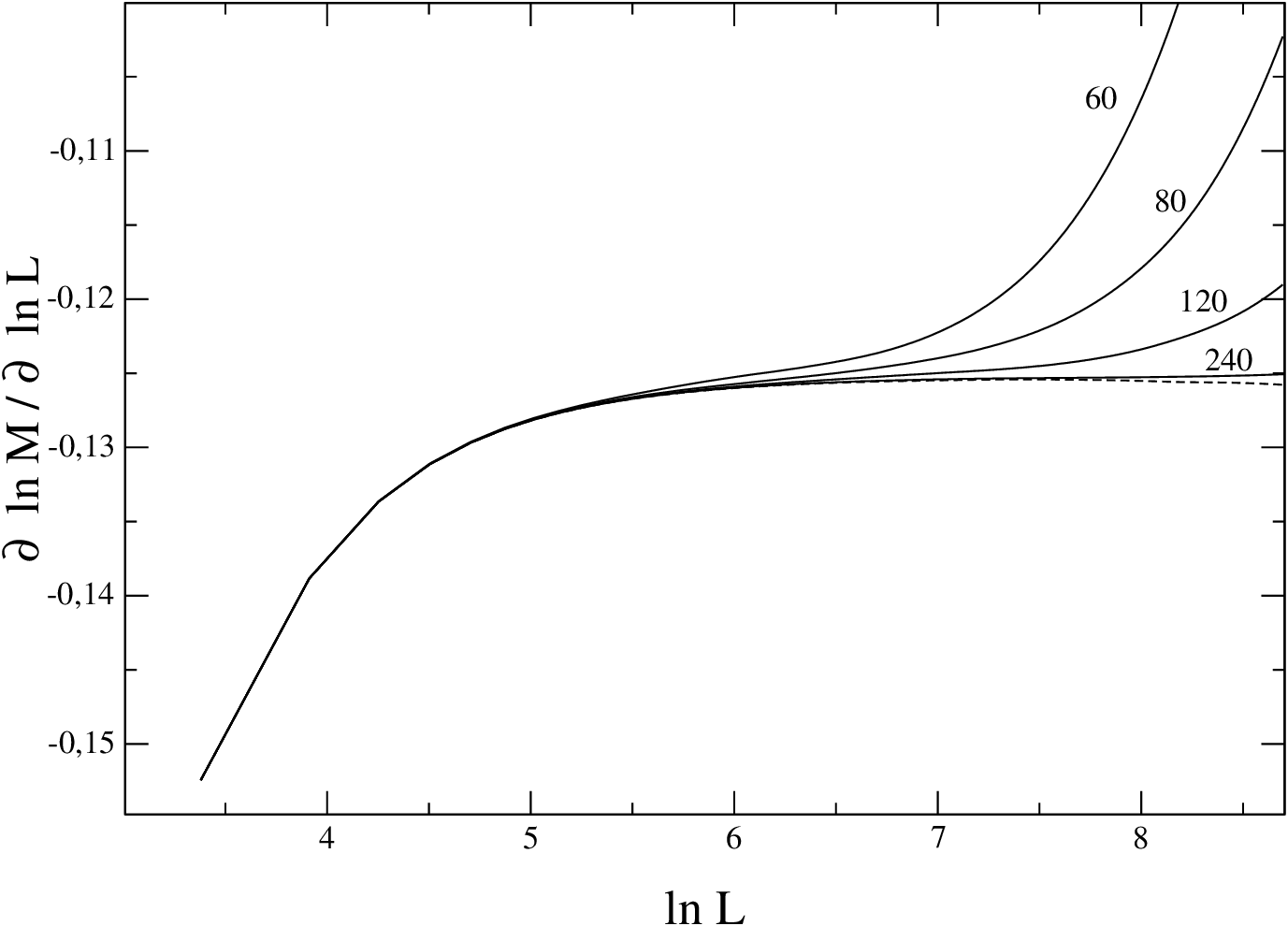} 
\caption{
$\partial \ln M/\partial \ln L$ plotted versus $\ln L$ for $\rho
=0.1$, $J=0.397726$. The solid curves are for  $m=60,80,120,240$. 
The dashed-line
curve is the result of an extrapolation to $m\to\infty$.
}
\label{fig5}
\end{figure}

\begin{figure}[htbp!]\centering  
\includegraphics*[width=7cm]{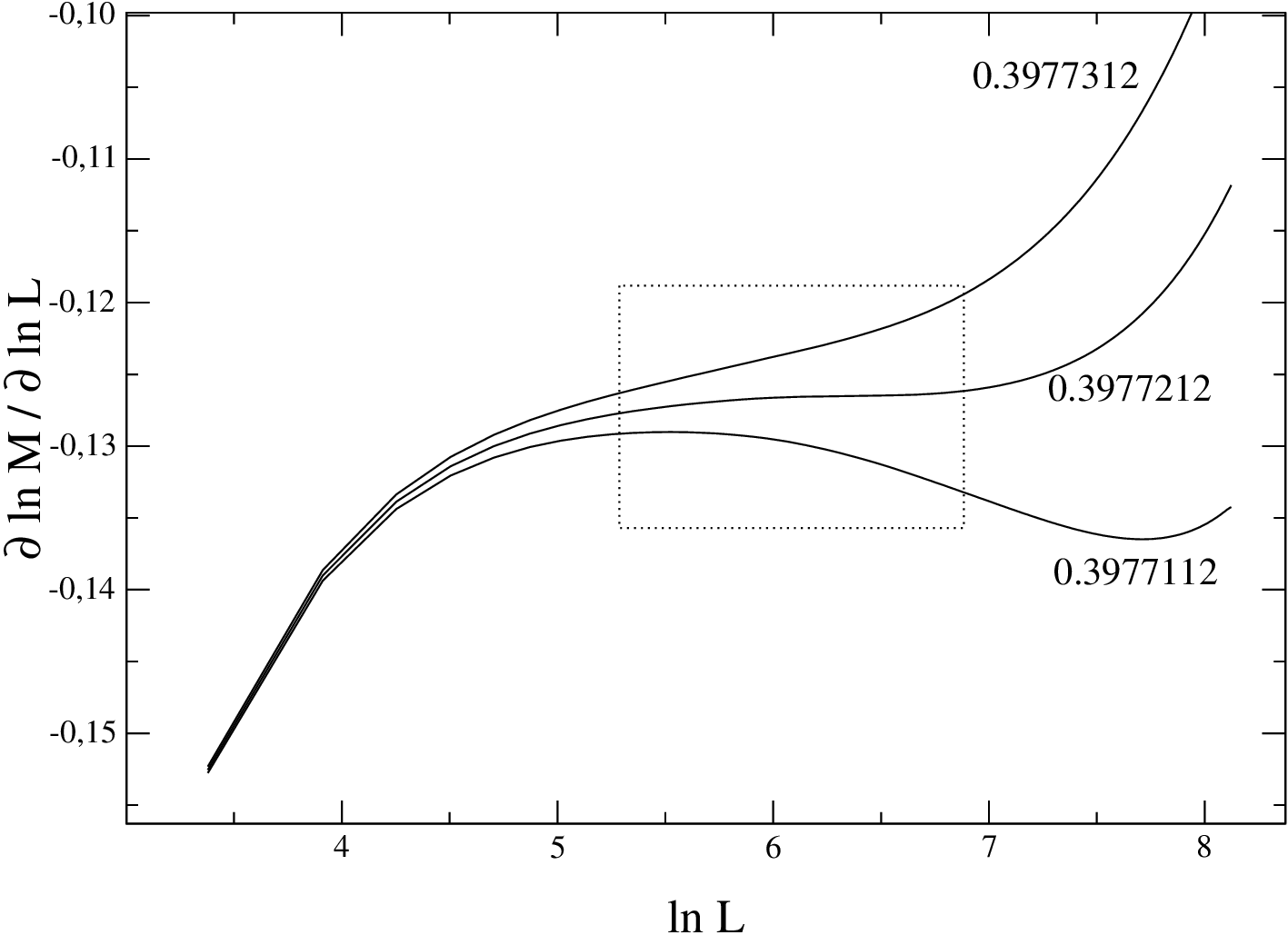} 
\caption{
$\partial \ln M/\partial \ln L$ plotted versus $\ln L$ for $\rho
=0.1$, $m=60$, for three different values of  $J$.
}
\label{fig6}
\end{figure}

Things are not so easy, however. Figure~\ref{fig5} 
shows the curves of $\partial \ln
M/\partial \ln L$ against $\ln L$ for $\rho =0.1,J=0.397726$ with different
renormalisation dimensions $m$. This $J$ value is quite near to $J_{c}$. 
For small $L$, there exists a region where the lattice size is too small and
finite-size scaling has not come into play. We will refer to this region as the
small-size region.  As $L$ increases, the number of states of the
system increases while the renormalisation dimension $m$ remains unchanged, and
we will suffer from the
``finite-$m$ effect'' in due time. To get a more complete knowledge of the finite-$m$
effect, let us look at Figure~\ref{fig6} 
where curves for different $J$ with $m=60$
are shown. $m=60$ is relatively small in our calculation and it has a more
drastic finite-$m$ effect. When $L$ is small, all the three curves shown in 
Figure~\ref{fig6} increase, which is an indication of the small-size effect. After a
certain value of $L$, one of the curves begins to bend down and the other two become
much flatter. Here the critical behaviour dominates. However, the finite-$m$
effect soon shows itself by raising the curves, which is also demonstrated
clearly in Figure~\ref{fig5}. As a result, only approximately the part inside the dotted-line box
resembles Figure~\ref{fig4}. It is hard to tell exactly where the finite-$m$ effect
begin to dominate without comparison with curves of higher $m$.

\begin{figure}[htbp!]\centering  
\includegraphics*[width=7cm]{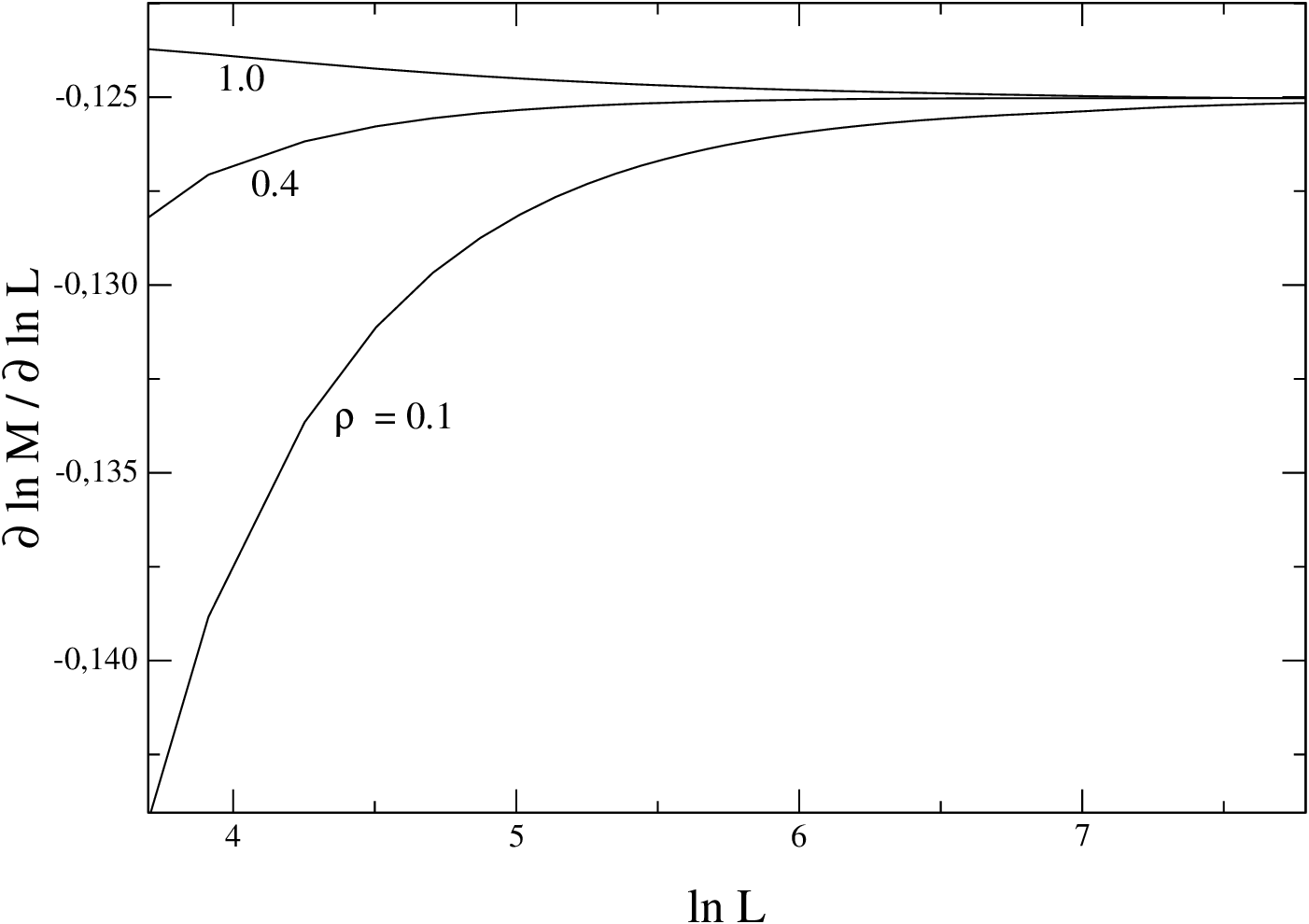} 
\caption{
$\partial \ln M/\partial \ln L$  plotted versus $\ln L$ for $J=J_{c}
$ with $\rho =1.0,0.4,0.1$ from above, respectively.
}
\label{fig9}
\end{figure}

Nevertheless, some useful information can be drawn from the analysis. If a
curve bends down for a period of $L$, one can judge that it corresponds to a 
$J$ that is smaller than $J_{c}$. In other words, if  a curve has a local
maximum, it corresponds to a $J$ with $J<J_{c}$. Thus the point now is to
find $J_{c}^{(60)}$, above which the corresponding curve has no local
maximum and below which it has. $J_{c}^{(60)}$ can be understood as a lower
limit for $J_{c}$. In general, one can obtain $J_{c}^{(m)}$ as an estimate
of $J_{c}$. The former will approach the latter from below as 
$m\rightarrow \infty $. Meanwhile the vertical value of the horizontal part
of the curve with $J=J_{c}^{(m)}$ gives the estimate for $(d-2+\eta )/2$.
To make the above discussion simpler, we have limited it only to the case
with similar behaviour as that of $\rho =0.1$. Later we will see in
Figure~\ref{fig9}, 
that the small-size effect may show itself differently as $\rho$ 
changes.
But similar analyses can still be developed for other cases.

In most of our calculations, we chose $m$ to be 60, 80, 120, and 240. For each 
$m$, we calculated the magnetisation with a set of $J$ values and then
determined the critical point using the method described above. The results
for $\rho =0.1$ are shown in Table~\ref{t1}. 
\begin{table}\centering
\begin{tabular}[t]{r|ll}
\hline
$m$ & \ \ $J_c^{(m)}$ & $(d-2+\eta)/2$ \\
\hline
60 & 0.3977212 & \hspace{0.4cm} 0.12629 \\
80  &0.3977230  &  \hspace{0.4cm}  0.12620 \\
120 & 0.3977252  & \hspace{0.4cm} 0.12548  \\
240 & 0.3977260  & \hspace{0.4cm} 0.12529 \\
\hline
$\infty$ & 0.3977264 & \hspace{0.4cm}  0.12520 \\
\hline
\end{tabular}
\caption{
The critical point and critical exponent $(d-2+\eta )/2$ at $\rho
=0.1$ estimated through first looking for the most flat curve of $\partial
\ln M/\partial \ln L\ \sim \ln L$ for certain $m$ then extrapolating to
infinite $m$.
}
\label{t1}
\end{table} 
The last row shows the extrapolated
values for $m=\infty $.

\begin{figure}[htbp!]\centering  
\includegraphics*[width=7cm]{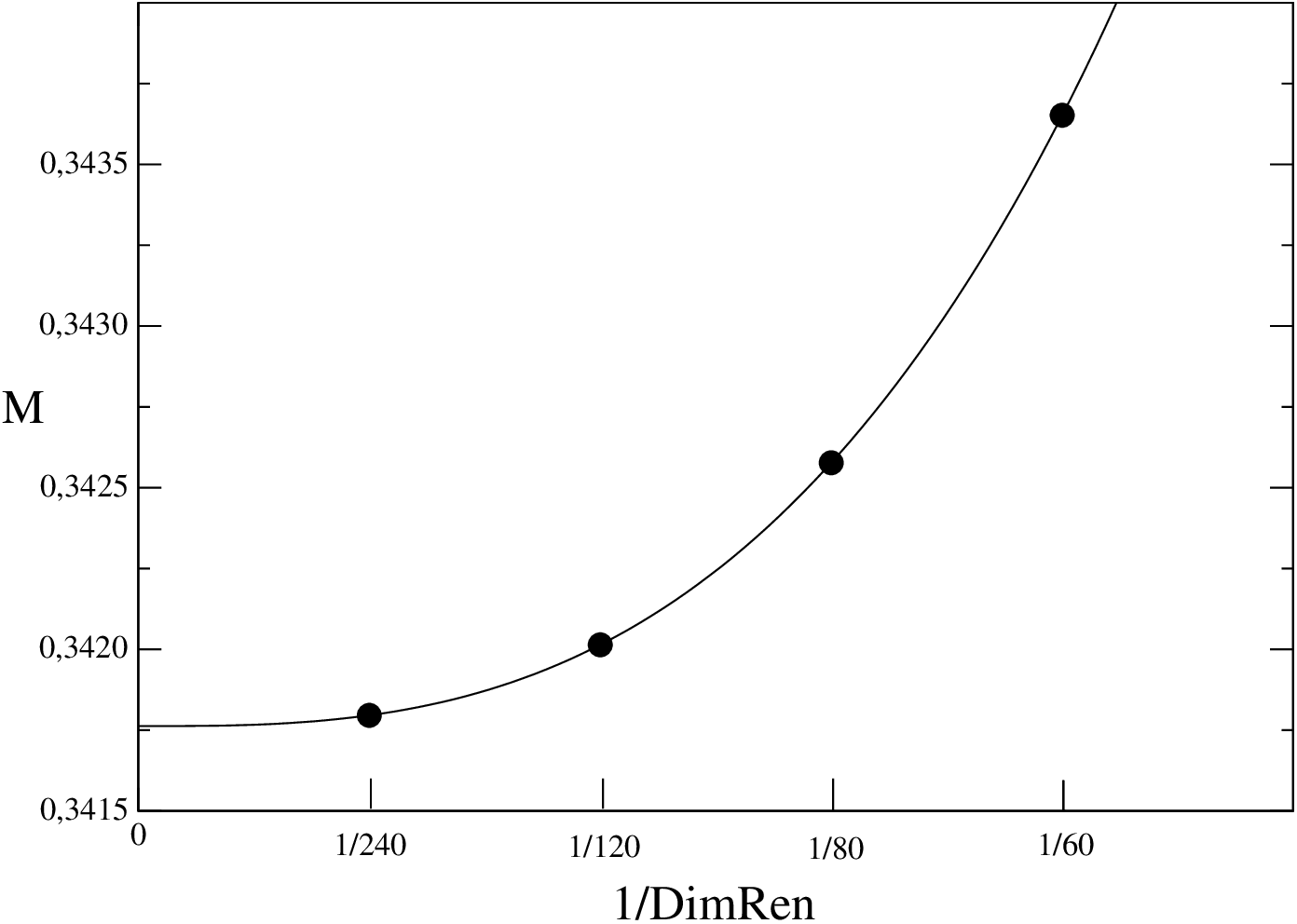} 
\caption{
Extrapolation from $m=60,80,120,240$ for $J=0.397726$, $\rho =0.1$
and $L=2001$.
}
\label{fig7}
\end{figure}

An alternate and finer way for dealing with the finite-$m$ problem is to extrapolate the data
to $m\to\infty$ first and then analyse the extrapolated data. The 
extrapolation should lead to reasonable and smooth curves
for $m=\infty$, and we assume the data converge monotonically  as 
$m\rightarrow \infty $. This defines our criterion for choosing the extrapolation
formula. 
We still take
as an example the case with $J=0.397726$, $\rho =0.1$, whose curves are plotted
in Figure~\ref{fig5}. For each value of lattice size $L$, we fit the points $\left(
1/m,M^{(m)}\left( L\right) \right) $, where $m=60,80,120,240$, with
\begin{equation}
y=a_{5}x^{5}+a_{4}x^{4}+a_{3}x^{3}+a_{0} \;. \label{fit}
\end{equation}
$a_{0}$ then is considered to be the value for $m=\infty $ 
(see Figure~\ref{fig7}).
Eq.(\ref{fit}) does not include  terms of $x^{1}$ and $x^{2}$ because 
including them violates the criteria above.
The extrapolated curve is also shown in Figure~\ref{fig5} as a
dashed line.

\begin{figure}[htbp!]\centering  
\includegraphics*[width=7cm]{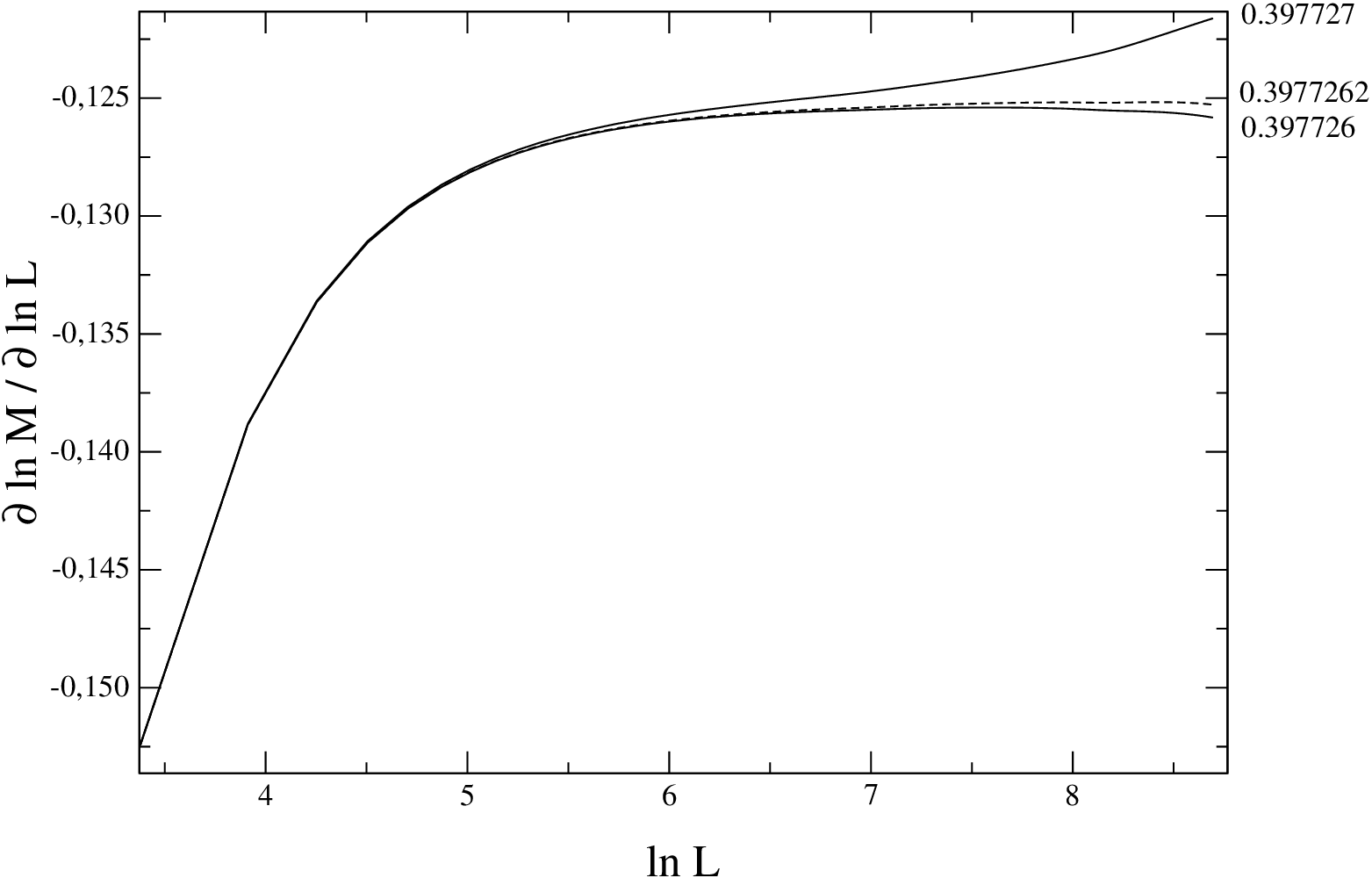} 
\caption{
$\partial \ln M/\partial \ln L$ plotted versus $\ln L$ for $\rho
=0.1$. The solid lines are the extropolated curves for $J=0.397726$ and
$J=0.397727$. The dashed line is obtained from an interpolation between the two
values of $J$. 
}
\label{fig8}
\end{figure}

Now that we obtained the curves for $m=\infty $ with different $J$, we can
search for the $J$ value whose $\partial \ln M/\partial \ln L \sim 
\ln L$ curve is closest to a horizontal line. This $J$ value is then the estimate
of $J_{c}$. In our calculation, we use interpolation technique to help
locate the ``best" $J$, as shown in Figure~\ref{fig8}.
 The solid lines are those
extrapolated to infinite $m$. We can obtain the magnetisation for any value of $J$ that is
between $J=0.397726$, for which the
curve bends down, and $J=0.397727$, where the line bends up, by linear
interpolation. We found $J_c$ to be  $0.3977262$ with  $(d-2+ \eta)/2=0.12507$ .

The small-size effect for different ratios is presented in Figure~\ref{fig9}. For small
$ \rho $, this effect is big. In the case of $ \rho =0.1$, for instance, the
small-size effect is visible with $L$ up to 800. However, it seems that we
overcome the small-size effect with $L=2000$.

Having $J_c$ in hand, we can determine another critical exponent $\nu$ through the
finite-size scaling form of the energy density
\begin{equation}
E(L)-E_\infty \sim L^{-(d-{\frac{1}{\nu }})}\;.
\end{equation}

The critical point $J_{c}$, critical exponents $(d-2+\eta )/2$ , and
$d-{\frac{1}{\nu}} $ for $\rho =0.1,0.4$, and $1.0$ are collected in
Table~\ref{t2}.
\begin{table}[ttbp!]\centering
\begin{tabular}[t]{r|lll}
\hline
$\rho$\ &  \ \ $J_c^{(\infty)}$ & $(d-2+\eta)/2$  & \ $d-\frac{1}{\nu}$\\
\hline
0.1 & 0.3977262 & \hspace{0.4cm} 0.12507 & 1.0014 \\
0.4 & 0.3541412 & \hspace{0.4cm} 0.12503 & 1.0001 \\
1.0 & 0.3117577 & \hspace{0.4cm} 0.12502 & 0.9996 \\
\hline
\end{tabular}
\caption{
Critical point $J_{c}$, critical exponents 
for $\rho =0.1,0.4$, and $1.0$, estimated through first extrapolating
the curve of $\partial \ln M/\partial \ln L \sim \ln L$ for infinite 
$m$ then looking for the most flat curve.
}
\label{t2}
\end{table}

We also estimated the shift exponent by taking  $\rho =0.005$ and $0.01$. The
critical points are found to be $J_{c}=0.432507$ and $0.428593$ respectively,
which lead to a value $1.773$ for the shift
exponent that consists with the theoretical prediction $1.75$ with $1.3\%$
deviation.

\vspace{1cm}

\paragraph{\bf Discussion and conclusions}
Our numerical results suggest that the TLIM model is in the Ising universality
class. That is, for a finite interlayer coupling, the critical behaviour of
the TLIM is the same as that of the one-layer system. Both  $<s>$ and  $<\sigma >$
are order parameters in the transition. Our observations  strongly exclude a
phase transition for the electric order, except at the decoupling point $\rho=0$. 

\begin{figure}[htbp!]\centering  
\includegraphics*[width=7cm]{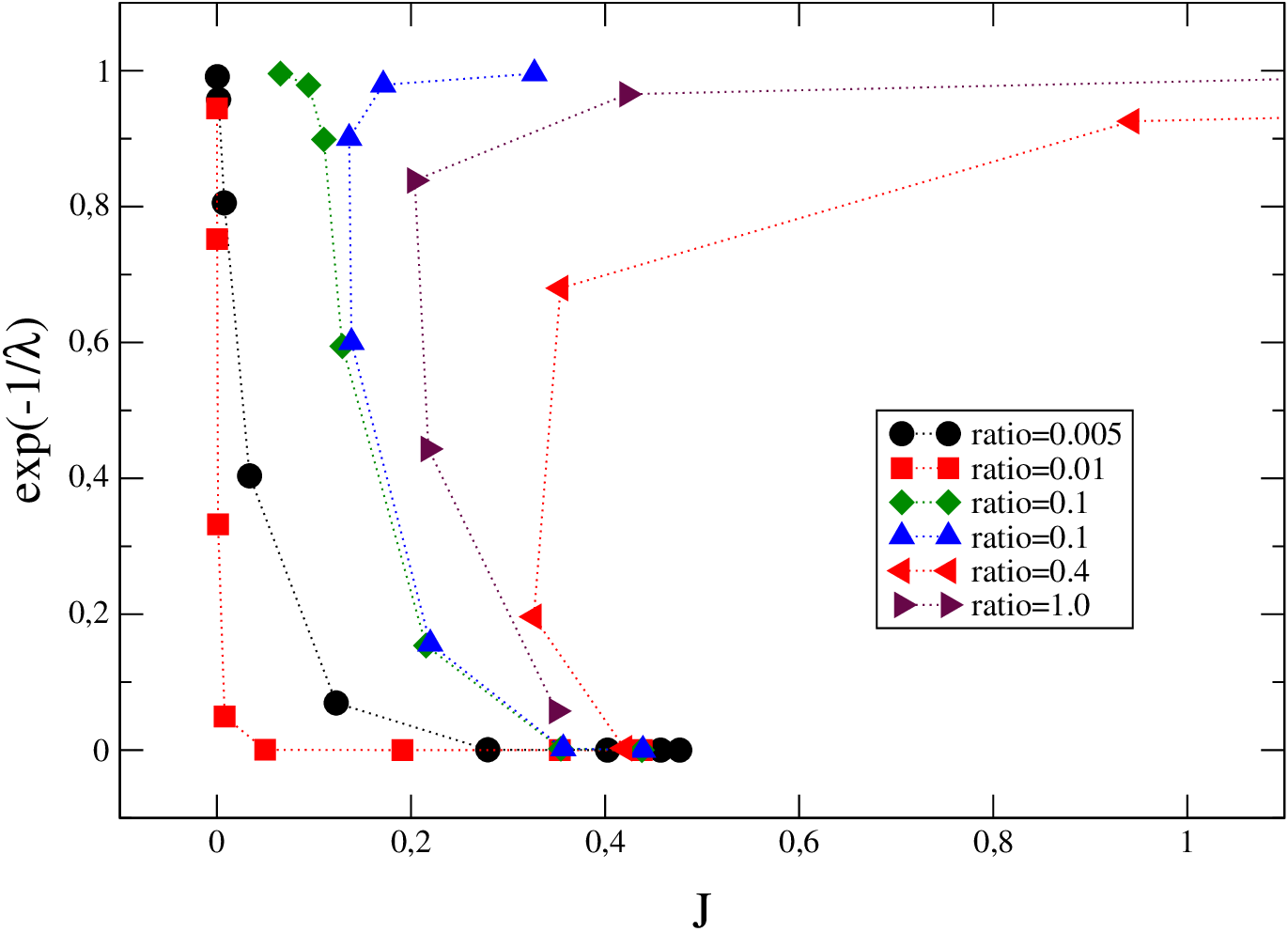} 
\caption{
The RG trajectories mapped to the TLIM parameter space are plotted.
The vertical axis is $\exp(-1/\lambda)$ with $ \lambda=\rho J$. Corresponding to
the curves in the order of the legend, the starting couplings are $J=0.477$,
$0.438$, $0.438$, $0.439$, $0.42$, and $0.35$, respectively.
}
\label{fig11}
\end{figure}

The  critical line of the TLIM is in fact not a fixed line since the
interlayer coupling is a relevant interaction which should flow to infinity
in renormalisation group transformations. Then the TLIM resembles the
one-layer Ising model in the long-range critical behaviour. 
To confirm this picture,
we also carry out the real-space renormalizational group transformation to
the model. The lattice is divided into $3\times 3$ blocks. All nearest
neighbour interactions are kept. The RG trajactories are plotted in 
Figure~\ref{fig11}. 
Two attractive fix points both having infinite $\rho $ are observed. 
They are the ordered and disordered fix point, respectively.
The
critical line seperates two attractive fix points and obviously it is not a
fix line.

However, as the
interlayer coupling is small, it needs a large scale transformation to reach
the infinite fixed points, so the model has extraordinary large finite-size
effects as we have observed in the numerical results.

The decoupling limit shows singularities due to the existence of the severe
competition between the unstable decoupling fixed point and the stable
infinite interlayer coupling fixed points which, however, are far away. This
supported in some sense by the unusual sensitivity of the shift exponent on
the couplings \cite{Hori97, Lipo98} and the strange layer symmetry breaking
in the effective state at the decoupling limit \cite{Ange97}. Our estimate
for the shift exponent is consistent with the theoretical prediction.

The unusual small-size effects will have important consequences on the
local properties of the TLIM\@. It is manifested in the short-time critical
behaviour that we recently observed in Monte Carlo simulations which will be
reported elsewhere \cite{Li00}.

\vspace{0.5cm}

\begin{acknowledgements}
The authors thank B. Zheng  for helpful
discussions.
\end{acknowledgements}

\end{document}